# Method of estimation of turbulence characteristic scales

V.A. Kulikov [1*], M.S. Andreeva [2], A.V. Koryabin [3], V.I. Shmalhausen [2]

[1] *A.M.Obukhov Institute of Atmospheric Physics Russian Academy of Sciences*

[2] *Chair of General Physics and Wave Processes Department of Physics, M.V.Lomonosov Moscow State University, Moscow 119991, Russia.*

[3] *International Laser Center, M.V.Lomonosov Moscow State University, Moscow 119991, Russia.*

*Corresponding author: vkulik@mail.ru

**Abstract**: Here we propose an optical method that use phase data of a laser beam obtained from Shack-Hartmann sensor to estimate both inner and outer scales of turbulence. The method is based on the sequential analysis of normalized correlation functions of Zernike coefficients. It allows excluding the $C_n^2$ value from the analysis and reduces the solution of a two-parameter problem to sequential solution of two single-parameter problems. The method has been applied to analyze the results of measurements of the laser beam that propagated through a water cell with induced turbulence and yielded estimates for outer and inner scales.

## 1. Introduction

Turbulence is one of the key factors responsible for light beam distortions during its propagation through randomly inhomogeneous medium such as the atmosphere. Many common methods of turbulence studies are based on the phase or amplitude analysis of the light wave that has passed through turbulent medium. In adaptive optics a decomposition of phase into a set of Zernike polynomials [1] is used to analyze phase distortions. On this way one can analyze correlation functions of phase decomposition coefficients [2, 3]. This approach allows exclusion of insignificant modes from the analysis. Expressions for correlation functions of Zernike coefficients for Kolmogorov model of turbulence were given in [4]. More complex models consider inner [5] and outer [6] scales of turbulence. Effects of these scales were investigated in [7, 8, 9 and 10]. The role of the outer scale in correlation functions of phase decomposition into Zernike polynomials has been subjected to detailed analysis in [7, 8]. The outer scale of atmospheric turbulence estimates given in different studies vary significantly. The $L_0$ value obtained in [11] by non-optical means occurred to be <5 meters. In [7] authors give the $L_0$ value between 16 m and 80 m based upon data from Keck telescope while [12] suggest 25 cm.

Measurements of the inner scale are discussed in [13,14]. A method for inner scale estimation that is proposed in [12] requires usage of two parallel beams [12, 15]. Studies mentioned give estimates of the inner scale value about 1 mm or about 10 mm in the atmosphere. Both characteristic scales are important for the description of turbulence and understanding of its optical effects.

To describe turbulence we used a von Karman-Tatarski model. Theoretical analysis shows that the dependence of Zernike coefficients correlation functions on changes in characteristic turbulence scales is not uniform. We supposed to use lower modes in the estimation of the outer scale. As simultaneous calculation of both outer and inner scales from the same functions gives ambiguous results, we suggest using correlations of higher order modes for estimation of the inner scale which, in turn, allows estimation of the outer scale from correlations of lower modes. Thus we reduced the two-parameter problem ($L_0$, $l_m$), which does not have unambiguous solution to sequential solution of two single-parameter problems. We also propose a method for estimation of turbulence characteristics that is based on above-mentioned correlations. This method has been applied for analysis of the experimental data from turbulence modeling in a water cell with a laser beam passing through it.

Early attempts to use temperature gradient in water are described in [16, 17], later this model has been used in [18]. We used an improved version of this method to estimate inner and outer scales of turbulence occurring in a water cell under various temperature conditions.

Estimates of turbulence scales in water may be applied to build a theory of underwater image transfer [19] or to studies of oceanic turbulence [20].

The method proposed includes the analysis of correlations for Zernike modes of at least the first order (to estimate $L_0$) and third order (to estimate $l_m$), that is 6 curves total. As estimates of the inner scale derived from third-order curves are very close, any of the four third-order correlations may be used in the case of isotropic turbulence, while the outer scale may be estimated from any of the first-order curves. Thus, a rough estimation of scales requires correlation functions for two modes.

The experiment involved a single wide laser beam and a wavefront sensor with two virtual subapertures allocated in its receiving aperture.

## 2. Method of estimation of turbulence scales

A relatively large value of inner scale in water turbulence [18] allows us assume it's important role. We make calculation of the Zernike polynomials correlation functions for two separated apertures analogously to [7], but we consider more complex form of spectrum accounting both inner and outer turbulence scales. The analyses of obtained data let us suggest the method of sequential estimation of the turbulence scales.

Expansion of phase of wave front on a receiving aperture on Zernike polynomials to be

$$\varphi(R\rho,\theta) = a_j Z_j(\rho,\theta), \quad (1)$$

where $j$ is number of mode, the Zernike polynomials $Z_j(\rho,\theta)$ are defined in the same manner as that in Noll [1].

We consider the phase in two separated apertures

$$\varphi_1 = \varphi(R\rho_1,\theta), \quad (2)$$
$$\varphi_2 = \varphi(R(\rho_2 + 2S),\theta), \quad (3)$$

where S is a vector connecting the centers of two coplanar receiving apertures. Then the coefficients $a_{1j}$ and $a_{2j}$ are represented as

$$a_{ij} = \int \varphi_i Z(\rho_i) W(\rho_i) d\rho_i, \quad (4)$$

where $i$ is the index marking the first and the second apertures, $W(\rho)$ is the window function defined by

$$W(\rho) = \begin{cases} 1/\pi & |\rho| \leq 1 \\ 0 & |\rho| > 1 \end{cases}, \quad (5)$$

and the integration is performed over infinity. The correlation of the coefficients is then given by

$$\langle a_{1j} a_{2j}^*(S) \rangle = \int d\rho_1 \int d\rho_2 C[R(\rho_2 + 2S - \rho_1)] \times Z_{1j}(\rho_1) Z_{2j}^*(\rho_2) W^*(\rho_2), \quad (6)$$

where $C$ is the phase correlation function given as

$$C[R(\rho_2 + 2S - \rho_1)] = \langle \varphi_1 \varphi_2^* \rangle.$$

Using Parseval's equation, the Wiener–Khintchine theorem, and the convolution theorem, we can write Eq. (6) in Fourier space:

$$\langle a_{1j} a_{2j}^*(S) \rangle = \int d\vec{k} \frac{1}{R^2} exp[-2\pi i \vec{k}(2S)] \Phi_S(\frac{k}{R}) \tilde{Z}_{1j}(\vec{k}) \tilde{Z}_{2j}^*(\vec{k}), \quad (7)$$

where $\tilde{Z}_{ij}$ is the Fourier transform $Z_{ij}(\rho_i) W(\rho_i)$.

For simplicity we suppose that aperture separation is realized along axis Y. Two-dimensional spectral density of the phase fluctuation in the von Karman-Tatarskii model is

$$\Phi(k) = \frac{A \exp\{-(k/k_m)^2\}}{k(k^2 + k_0^2)^{11/6}}, \quad (8)$$

where $A = 0.0096(2\pi/\lambda)^2 \int_0^\infty C_n^2(z) dz$, $C_n^2$ is structural constant of refractive index, $\lambda$ is the wavelength of the propagating light, $k$ is wave vector, $k_0 = 1/L_0$, $L_0$ is the outer scale of turbulence; $l_m = \alpha/k_m$, $l_m$ is the inner scale of turbulence, $\alpha = 5.92$.

Thereby the dependence of Zernike modes correlation functions on the distance and changes in characteristic scales can be described by expression

$$\langle a_{1j} a_{2j}(S) \rangle = A(n+1) \int_0^\infty e^{-\frac{k^2}{k_m^2}}$$
$$\times (J_0(2Sk) + pJ_{2m}(2Sk)) \frac{J_{n+1}^2}{k(k^2 + k_0^2)^{11/6}} dk, \quad (9)$$

where $n$ is the radial order of the $j$-th Zernike polynomial, J is the Bessel functions of an appropriate order; $p=0,-1,+1$ depending on Zernike mode number.

In our analysis we used the normalized value of

$$C = \frac{\langle a_{1j} a_{2j} \rangle}{\langle a_1 \rangle^2}, \quad (10)$$

in assumption of $\langle a_{1j} \rangle^2 = \langle a_{2j} \rangle^2$. This normalization eliminates the analysis of phase fluctuations dispersion which corresponds to the turbulence intensity of the Kolmogorov model. The normalized correlation function may be considered to depend only on characteristic scales of turbulence.

Attempting to estimate characteristic scales of turbulence from one of the lower order modes we gained an ambiguous result since one experimental curve may be approximated with a set of different pairs of values for inner and outer scales. To eliminate this problem we analyzed relationships between characteristic turbulence scales and correlation functions of first three orders of Zernike polynomials.

Figure 1 displays partial results of this analysis, one mode for each radial order. Such kind of representation is acceptable since responses of correlations for modes of any order to changes in characteristic scales are similar. Note that correlations for first-order modes depend on both scales of turbulence.

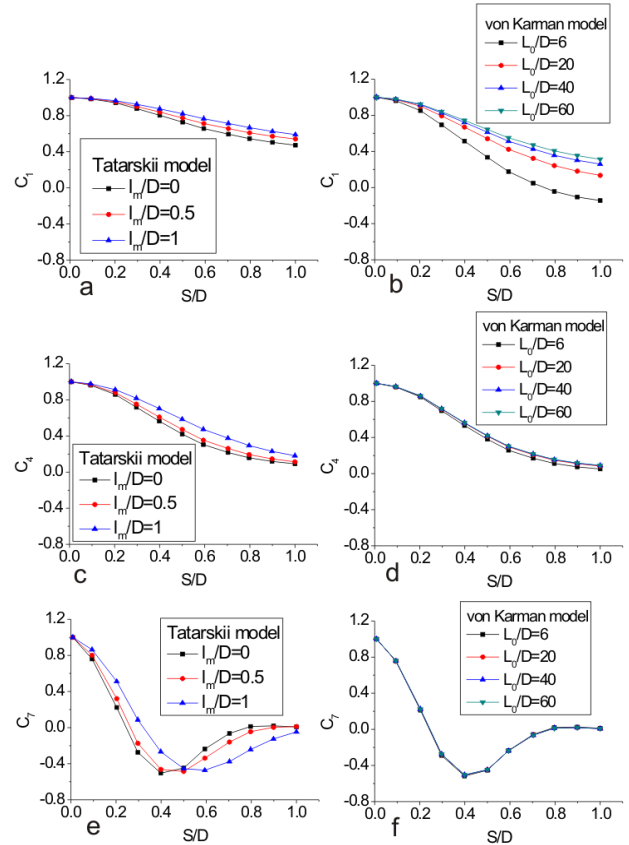

Figure 1. The dependence of correlation functions on the S/D parameter at various values of $L_0/D$ in von Karman model: a) for the 1st mode; b) 4th mode; c) 7th mode; and at various values of $l_m/D$ in Tatarski model: d) 1st mode; e) 4th mode; f) 7th mode.

The behavior of second-order curves which are represented by j=4 shows significant dependence from the outer scale changes only if $L_0/D<10$. It can be seen that the sensitivity of second-order curves to outer scale changes is substantially lower then that of the first order curves. Dependence of correlations for second-order modes on the inner scale changes is not quite linear and becomes obvious at $l_m$ values greater than 0.5.

Dependence of correlations for third-order modes on the outer scale changes is neglectable in the considered range of $L_0/D$ values, while their dependence on the inner scale changes can be seen easily. Note that taking the inner scale into account results in a rightward shift of the predicted minimum of the correlation function for the 7-th mode.

The unequal impact that scales have on different correlation functions is due to the fact that outer and inner scales affect various parts of the spectrum in different ways. The decomposition of phase aberrations into Zernike modes results in spatial filtration that allows discrimination of contributions from large-scale and small-scale perturbations.

In order to get optimum estimates for turbulence scales it is reasonable to determine the inner scale value first, using correlations of third-order modes as this scale is the only factor affecting them. After that the outer scale value can be determined from correlations of lower order modes given the inner scale is already known. Correlations of first-order modes are most suitable for that as their dependence from the outer scale of turbulence is the most pronounced.

So, the method we propose for estimation of characteristic turbulence scales is based upon sequential analysis of normalized correlation functions of Zernike coefficients. The method allows elimination of the $C_n^2$ value from the three-parameter problem with $L_0$, $l_m$, $C_n^2$ parameters and reduces solution of the resulting two-parameter problem to sequential solution of two single-parameter problems. The first of these problems is estimation of the inner turbulence scale from correlation functions for Zernike modes of third and higher orders, which depend only on $l_m$ and the second one is estimation of the outer scale from first-order correlation functions which depend on both $L_0$ and $l_m$. The method has been applied to a collimated laser beam passing through a water cell.

## 3. Experimental design

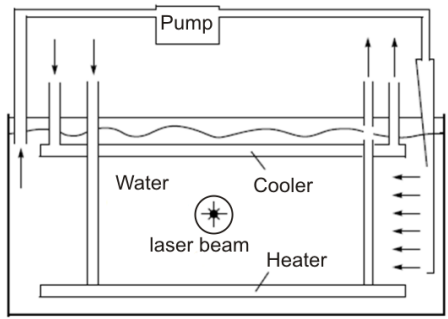

Figure 2. Design of the experimental assembly

Figure 2 shows the scheme of a liquid cell where turbulence has been induced. It was a rectangular cell of optical glass, 34 cm in length and 21 cm width, filled with water.

In nature turbulence is caused by temperature gradients (e.g. near heated surface of ground or ocean), winds and currents. In our experiments turbulence has been induced with a vertical temperature gradient created between lower heating plate and upper cooling plate. Distance between heater and cooler was 10 cm. Both heater and cooler were flat hollow heat exchangers with hot and cold water circulating through them, respectively, maintaining temperature difference constant.

To simulate turbulence driftage caused by a wind we equipped the cell with outlet and intake nozzle connected to a circulation pump with pipes of reinforced plastic. The flow rate was maintained at 0.8-1.2 cm/sec. To control the experimental assembly parameters we used a PC equipped with ADC card connected to 4 semiconductor temperature sensors. These sensors measured water temperature near intake and outlet nozzles ($t_1$ and $t_2$), near the cooling plate ($t_3$) and within the liquid layer in the cell ($t_4$).

A 30 mm wide collimated laser beam was directed to the cell so that it was orthogonal both to the temperature gradient and the water flow (perpendicularly to the plane of Figure 2). The wavefront of laser beam that passed the cell was being detected with a Shack-Hartmann type sensor with 30-mm input aperture. The sensor contained 1500 subapertures that provided spatial resolution about 0.75 mm. The maximum measurement rate was about 56 Hz. The control software was able to reconstruct the phase profile of the beam at the receiving aperture from local slopes measured at subapertures and to present it as Zernike polynomial expansion. Polynomials up to third radial order, nine polynomials total, were involved in our experiments.

For measurement of phase correlations two virtual circular apertures, both 15 mm in diameter, were allocated in the reconstructed distribution.

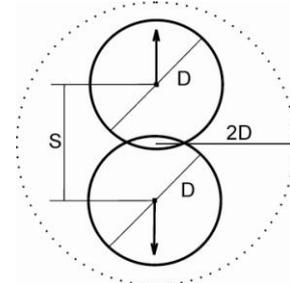

Figure 3. The direction of aperture shifting (*y* axis)

Relative position of these apertures to each other could be shifted for a distance S in *y* direction as it is shown at Figure 3 (along the temperature gradient). Correlation functions of phase distortions were then calculated based on phase data from these apertures.

## 4. Results

### Thermophysical calculations

Measurements have been performed at the following values of temperature difference: $\Delta t=0$, 10, 15, 20, 25°C without the flow and $\Delta t=0$, 10, 15°C with the flow turned on. Measurements at zero temperature difference were made to control the minimum noise. Without heating ($\Delta t=0$°C) non-normalized values of mode correlations are more than two orders of magnitude smaller than with it, thus they can be neglected.

To evaluate processes in the cell we have calculated such parameters as Rayleigh (Ra) number and Prandtl (Pr) number

$$Ra = \frac{g\alpha\Delta T l^3}{\nu\chi}, \quad Pr = \frac{\alpha}{\nu},$$

where g is the gravity factor, $l$ is a characteristic length ($l$=10 cm), $\Delta T$ is temperature difference, $\nu$ is kinematic viscosity of the liquid, $\chi$ is thermal diffusivity, $\alpha$ is thermal expansion factor.

Table 1

| $\Delta t$, °C | 10 | 15 | 20 | 25 |
|---|---|---|---|---|
| Ra | $2.36*10^8$ | $4.13*10^8$ | $5.71*10^8$ | $7.14*10^8$ |
| Pr | 6.27 | 5.17 | 5.14 | 4.76 |

Table 1 contains calculated values of Rayleigh and Prandtl numbers for all studied modes of turbulence. These values mean strong turbulence at all of the values of temperature difference we studied.

Note that at flow rates V=0.8-1.2 cm/sec the Reynolds number value that corresponds to cell parameters ($Re = \frac{VL}{\nu}$, L is a characteristic flow size, L=10 cm) Re=(1-1.5)*$10^3$ is below the critical value so the flow should not be turbulent.

**Estimation of turbulence parameters**

While analyzing the experimental results we checked turbulence for isotropy, the procedure is described in detail in Appendix A. Briefly, the isotropic theory states that some Zernike modes (e.g. j=1 and j=2, j=7 and j=8) fit each other when the direction of aperture shifting changes from y to x. We show that third-order modes meet this requirement in all modes of turbulence we studied. First-order modes demonstrate such fit only at $\Delta t$=10°C and are very close to that at $\Delta t$=15°C. With further raise of temperature or introduction of the flow the isotropy becomes disturbed. We suggest the anisotropy of large-scale fluctuations to be due to the effect of cell walls.

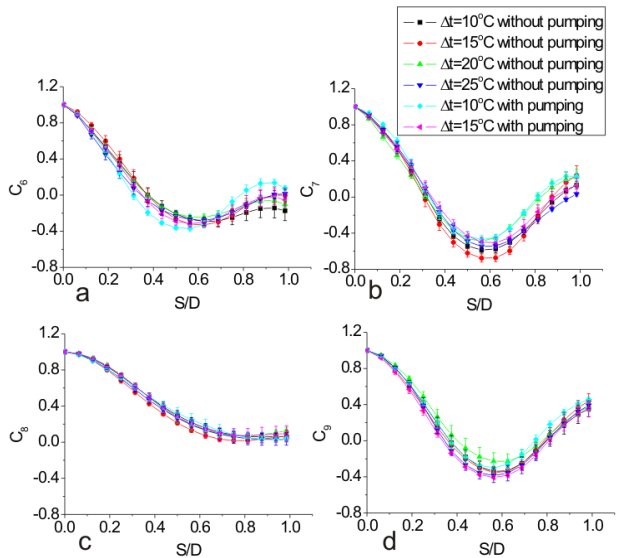

Figure 4. Comparison of experimental correlations for Zernike modes of collimated beam obtained at different values of aperture shift S with aperture size D in presence or absence of a flow at temperature difference values $\Delta t$=10, 15, 20, 25°C. a) j=6; b) j=7; c) j=8; d) j=9

Figure 4 displays all correlation functions for third-order modes. Note that at different temperature regimens they fit each other within the error. This insensitivity to changes in temperature and flow suggests that in all regimens we studied characteristics that correspond to small-scale turbulence are very similar.

Approximation of these data with theoretical curves produces nearly equal lm values (lm=15±5 mm) at any temperature difference. It can be clearly seen that presence or absence of the flow does not affect third-order correlation curves and the inner scale estimate. So, the inner scale does not depend on temperature and wind fluctuations. The estimation procedure for the inner scale is described in Appendix B along with some problems that occur.

An example of the method application is shown at Figure 5 for temperature difference $\Delta t$=10°C and zero flow. Only two of four third-order modes are presented. The full set of correlations for third-order Zernike modes at $\Delta t$=10°C is presented and discussed in Appendix B. It should be emphasized that correlation functions presented at Fig. 5ab can't be described using Kolmogorov and von Karman models, as they are insensitive to the outer scale changes up to the aperture diameter.

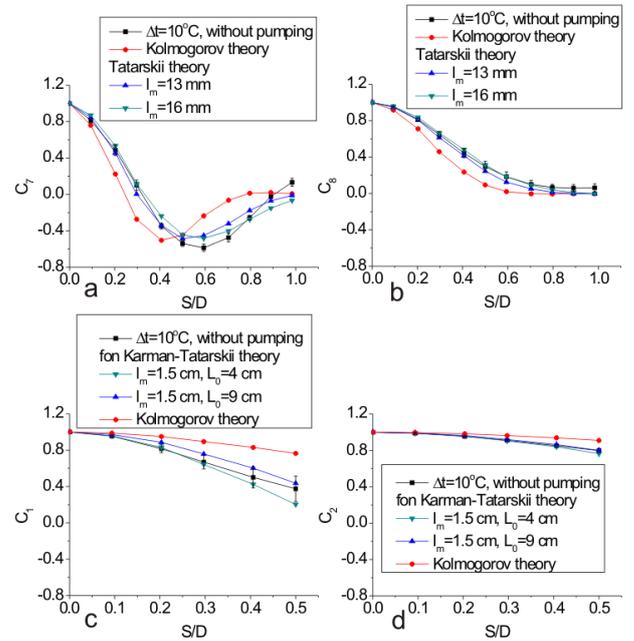

Figure 5. Sequential estimation of characteristic scales of turbulence. a) Estimation of $l_m$ from $K_7$; b) Estimation of $l_m$ from $K_8$; c) Estimation of $L_0$ from $K_1$; d) Estimation of $L_0$ from $K_2$

In isotropic conditions at $\Delta t$=10°C the method yields values of $L_0$= 4-9 cm and $l_m$=1-2 cm. Estimates of characteristic scales for other values of temperature difference were obtained in the same way. Resulting estimates are given in Table 2 below.

Table 2

| $\Delta t$, °C | pumping | $l_m$, cm | along Y; $L_0$, cm | along X $L_0$, cm |
|---|---|---|---|---|
| 10 | no | 1-2 | 4-9 | 6-10 |
| 15 | no | 1-2 | 1.5-2.5 | 2-3 |
| 20 | no | 1-2 | 12-20 | 3-6 |
| 25 | no | 1-2 | 6-12 | 3-5 |
| 10 | yes | 1-2 | 28-33 | 8-9 |
| 15 | yes | 1-2 | 7-10 | 6-9 |

This table shows clearly that as the temperature difference increases estimates of the outer scale derived from $K_1$ and $K_2$ functions no longer match. This may be caused by the rise of turbulence anisotropy. Note that flow leads to the increase of the outer scale. Estimates of the inner scale don't vary along with changes in thermal parameters and flow due to isotropy of small-scale fluctuation statistics.

## 5. Conclusion

The method developed allows estimate of both inner and outer scales of turbulence in a water cell in all situations we considered. Obtained estimates for the inner scale $l_0$=1-2 cm are close to those made in [18].

Theoretical analysis shows that outer scale almost does not affect correlation functions for Zernike modes higher than of second radial order when it is larger than the size of receiving aperture. Taking the outer scale into account is necessary for correct interpretation of experimental correlations obtained for first-order Zernike modes only when $L_0/D<50$.

The outer scale appeared to be sensitive to changes in temperature difference and varied in a range from 2 to 30 cm. This may be explained by change of the convection flows intensity that occurs in the cell when the temperature difference changes, as these flows determine the maximum size of turbulent irregularities. It also should be noted that introduction of the flow increases estimated value of the outer scale.

We also established that if the temperature difference is small $\Delta t=10^{\circ}C$; $15^{\circ}C$ and no flow is present, respective estimates of the outer scale from correlation functions $K_1$ and $K_2$ are equal. This means that turbulence is isotropic at these $\Delta t$ values. As the temperature difference increases to $\Delta t=20^{\circ}C$; $25^{\circ}C$ these estimates become different, with increase of the outer scale value that was estimated from $K_1$ (slope to $y$, direction of the temperature gradient).

Theoretical analysis makes evident that correlation functions undergo significant changes when the inner scale $l_m/D=0.5$ or higher. Taking account of the inner scale when it is comparable to the aperture size allows in most cases satisfactory description of experimental data. In our experiment the inner scale appeared not to depend on flow or changes in temperature difference.

### Appendix A. Anisotropy

It is known that in the case of isotropy correlation functions, which transit into each other if $x$ is changed for $y$ should coincide when the direction of aperture shifting is changed from y to x. Examples are correlation functions for modes with j=1,2 and j=7,8.

All modes of turbulence we studied were checked for this criterion of isotropy. Figures A1-A4 present experimental correlations for first-order Zernike modes, which, according to isotropic theory, must fit each other.

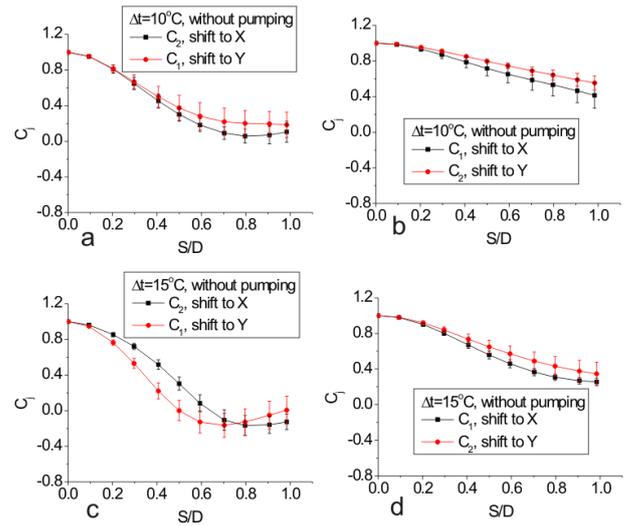

Figure A1. Correlation functions for Zernike modes (j=1, 2) with aperture shifting in $x$ direction (black curves) or $y$ direction (red curves) and no flow. a) $K_2$ ($X$ axis) and $K_1$ (Y axis), $\Delta t=10^{0}C$; b) $K_1$ ($X$ axis) and $K_2$ (Y axis), $\Delta t=10^{0}C$; c) $K_2$ ($X$ axis) and $K_1$ (Y axis), $\Delta t=15^{0}C$; d) $K_1$ ($X$ axis) and $K_2$ (Y axis), $\Delta t=15^{0}C$

It is evident that curves fit each other at $\Delta t=10^{\circ}C$ and no flow is present. At $\Delta t=15^{\circ}C$ without the flow curves are very close but already not identical as Fig.A1c shows.

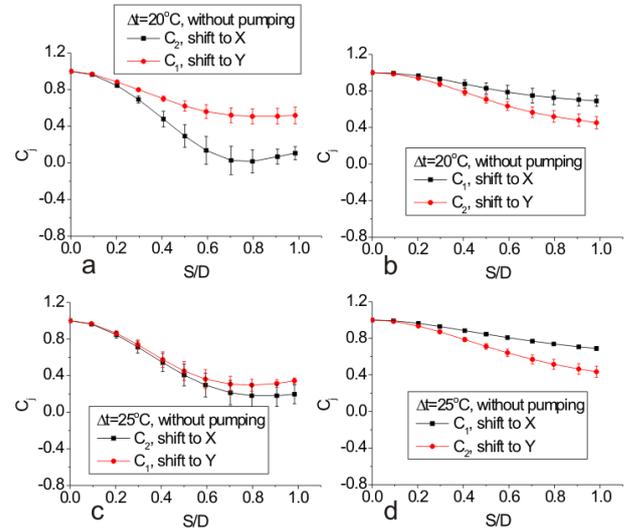

Figure A2. Correlation functions for Zernike modes (j=1, 2) with aperture shifting in $x$ direction (black curves) or $y$ direction (red curves) and no flow. a) $K_2$ ($X$ axis) and $K_1$ (Y axis), $\Delta t=20^{0}C$; b) $K_1$ ($X$ axis) and $K_2$ (Y axis), $\Delta t=20^{0}C$; c) $K_2$ ($X$ axis) and $K_1$ (Y axis), $\Delta t=25^{0}C$; d) $K_1$ ($X$ axis) and $K_2$ (Y axis), $\Delta t=25^{0}C$

Figure A2 displays first-order curves for $\Delta t=20$ and $25^{0}C$ with no flow. It is evident that the prognosis given by isotropic theory fails.

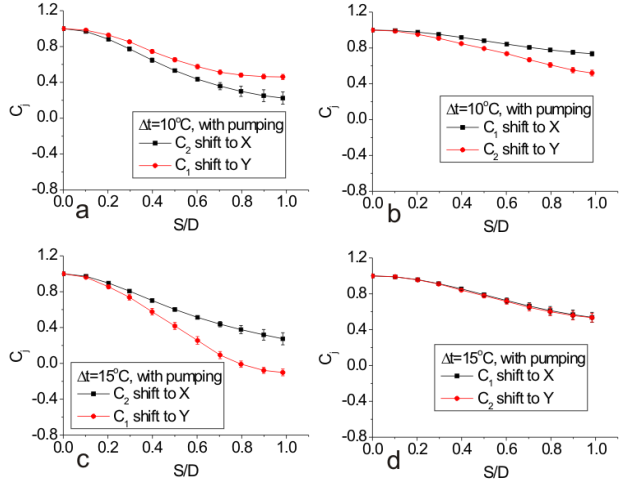

Figure A3. Correlation functions for Zernike modes (j=1, 2) with aperture shifting in *x* direction (black curves) or *y* direction (red curves) and in presence of a flow. a) $K_2$ (*X* axis) and $K_1$ (*Y* axis), $\Delta t=10^{0}C$; b) $K_1$ (*X* axis) and $K_2$ (*Y* axis), $\Delta t=10^{0}C$; c) $K_2$ (*X* axis) and $K_1$ (*Y* axis), $\Delta t=15^{0}C$; d) $K_1$ (*X* axis) and $K_2$ (*Y* axis), $\Delta t=15^{0}C$

Anisotropy can also be observed with the introduction of a flow at $\Delta t=10$ and $15^{o}C$ (Fig. A3).

We consider correlation functions for third-order Zernike modes which should fit according to the theory of isotropy. Experimental data for $\Delta t=10$, 15, 20 and $25^{0}C$ without a flow and for $\Delta t=10$ and $15^{0}C$ in the presence of a flow are analyzed. They are not presented as considered functions fit each other within the error. This suggests isotropy of a small-scale turbulence. A possible interpretation of this fact can be driftage of a small-scale turbulence by a flow without changing its statistical properties.

So the anisotropy is apparent only in correlations of lower modes that describe large-scale turbulence. It rises as the temperature difference increases or a flow appears. Correlations of higher Zernike modes that describe small-scale turbulence are in accordance with theory of isotropy.

**Appendix B. Problems with estimation of the inner scale**

Let's take a more close view to problems that rise while estimating of the inner scale. The maximum shift we used in approximation was S/D=0.5. Estimates of the inner scale that are discussed below may be applied to other turbulence modes that we investigated as correlations functions for them are nearly similar (Fig. B1). The correlation function for sixth mode can't be described using the theory we adopted. Introduction of a finite inner scale would not give any benefit as taking the inner scale into account can only lift the curve (i.e. increase the degree of correlation). This suggests that the turbulence we dealt with does not match any of the models used. Such behavior may be due to disagreement between the real turbulence spectra and model spectra. A small value of the inertial range that we obtained (it was also small in water cell experiments described in [18]) suggest usage of novel models for the description of water turbulence.

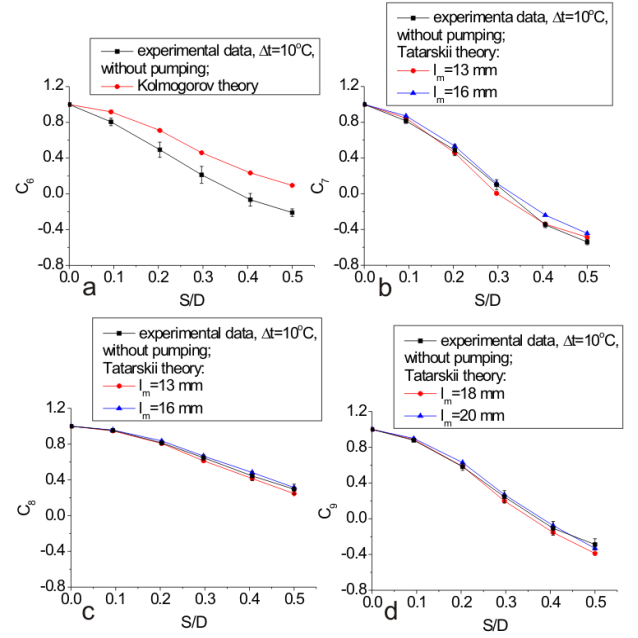

Figure B1. Experimental correlations for Zernike modes obtained at $\Delta t=10^{0}C$ with no flow, their approximations with Tatarski model with aperture of size D=1.5 cm shifting for distance S and estimated values of the inner scale $l_m$; a) j=6, $l_m$ not defined; b) j=7, $l_m$=13-16 mm; c) j=8, $l_m$=13-16 mm; d) j=9, $l_m$=18-20 mm

Taking the inner scale into account allows description of experimental data for most of the correlation functions within the approximation we chose. The $l_m$ values are nearly equal in correlations for modes j=7, 8, 9 at all temperature difference values we investigated. Taking account of the inner scale for first-order modes reduces estimated value for the outer scale. If no account of the inner scale is taken then the outer scale appears sometimes to be larger than the cell dimensions which is counterintuitive.

**6. Reference**

1. R.J. Noll, "Zernike polynomials and atmospheric turbulence," J. Opt. Soc. Am. **66,** 207–211 (1976).
2. G.C. Valley and S.M. Wandzura, "Spatial correlation of phase-expansion coefficients for propagation through atmospheric turbulence," J. Opt. Soc. Am. **69,** 712–717 (1979).
3. P.H. Hu, J. Stone, T.J. Stanley, "Application of Zernike polynomials to atmospheric propagation problems," J. Opt. Soc. Am. A **6**, 10, 1595–1608 (1989).
4. A.N. Kolmogorov, "Local structure of turbulence in an incompressible fluid at very high Reynolds number," Dokl. Akad. Nauk. USSR **30**, 299–303 (1941).
5. V. I. Tatarski, *Wave Propagation in a Turbulent Medium* (McGraw-Hill, New York, 1961).
6. T. von Kármán, "Progress in the statistical theory of turbulence," J. Mar. Res. **7**, 252-264 (1948). Reprinted in S. K. Friedlander and L. Topper, Turbulence (Interscience, 1961), 161-174.
7. N. Takato, I. Yamaguchi, "Spatial correlation of Zernike phase-expansion coefficients for atmospheric turbulence with finite outer scale," J. Opt. Soc. Am. A **12**. 5, 958-963 (1995).

8. D. M. Winker, "Effect of a finite outer scale on the Zernike decomposition of atmospheric optical turbulence," J. Opt. Soc. Am. A **8**, 10, 1568-1573 (1991).
9. A. Consortini, L. Ronchi, and E. Moroder, "Role of the outer scale of turbulence in atmospheric degradation of optical images," J. Opt. Soc. Am. **63**, 1246-1248 (1973).
10. G. C. Valley, "Long- and short-term Strehl ratios for turbulence with finite inner and outer scales," Appl. Opt. **18**, 984-987 (1979).
11. C.E. Coulman, J. Vernin, Y. Coqueugniot, and J.L. Caccia, "Outer scale of turbulence appropriate to modeling refractive index structure profiles," Appl. Opt. **27**, 1, 155-160 (1988).
12. A. Consortini and K.A. O'Donnell, "Measuring the inner scale of atmospheric turbulence by correlation of lateral displacements of thin parallel laser beams," Waves in Random Media, **3,** 85-92 (1993).
13. R.G. Frehlich, "Estimation of the parameters of the atmospheric turbulence spectrum using measurements of the spatial intensity covariance," Appl. Opt. **27**, 11, 2194-2198 (1988).
14. G.R. Ochs, Hill R.J., "Optical-scintillation method of measuring turbulence inner scale," Appl. Opt. **24**, 15, 2430-2432 (1985).
15. A. Consortini and K. O'Donnell, "Beam wandering of thin parallel beams through atmospheric turbulence," Waves in Random Media **3**, 11-28 (1991).
16. L.R. Bissonnette, "Atmospheric scintillation of optical and infrared waves: a laboratory simulation," Appl. Opt. **16**, 8, 2242-2251 (1977).
17. A.S. Gurvich, M.A. Kallistratova and F.É. Martvel', "An investigation of strong fluctuations of light intensity in a turbulent medium at a small wave parameter," Radiophysics and Quantum Electronics **20**, 7, 705-714 (1977).
18. Maccioni A., Dainty. J.C., "Measurement of thermally induced optical turbulence in a water cell," J. of Modern Opt. **44**, 6, 1111-1126 (1997).
19. W. Hou, "A simple underwater imaging model," Opt. Letters **34**, 17, 2688-2690 (2009).
20. D.J. Bogucki, J.A. Domaradzki, C. Anderson, H.W. Wijesekera, J.R.V. Zaneveld, C. Moore "Optical measurement of rates of dissipation of temperature variance due to oceanic turbulence," Opt. Express **15**, 12, 7224-7230 (2007).